\documentclass{article}
\usepackage{spconf,amsmath,graphicx}
\usepackage{hyperref}
\usepackage{amsfonts}
\usepackage{array,multirow,graphicx}
\usepackage{float}
\usepackage[font=footnotesize]{subcaption}

\usepackage{enumitem}
\setlist{nosep, leftmargin=14pt}

\usepackage{mwe} 

\def\L{{\cal L}}

\title{SegTransVAE: Hybrid CNN - Transformer with Regularization for medical image segmentation}
\name{Submission ID: 564}
\address{}

\name{Quan-Dung Pham$^{\star}$ \qquad Hai Nguyen-Truong$^{\star \dagger \ddagger}$ \qquad Nam Nguyen Phuong$^{\star}$ \thanks{Quan-Dung Pham and Hai Nguyen-Truong equally contributed.} \qquad Khoa N. A. Nguyen$^{\star \dagger \ddagger}$}
\address{$^{\star}$ VinBrain JSC., Vietnam 
        $^{\dagger}$ University of Science, Ho Chi Minh City, Vietnam \\ 
        $^{\ddagger}$ Vietnam National University, Ho Chi Minh City, Vietnam 
        ${\mathsection}$ Vin University, Vietnam}

\begin{document}
%
\maketitle
\begin{abstract}
Current research on deep learning for medical image segmentation exposes their limitations in learning either global semantic information or local contextual information. To tackle these issues, a novel network named SegTransVAE is proposed in this paper. SegTransVAE is built upon encoder-decoder architecture, exploiting transformer with the variational autoencoder (VAE) branch to the network to reconstruct the input images jointly with segmentation. To the best of our knowledge, this is the first method combining the success of CNN, transformer, and VAE. Evaluation on various recently introduced datasets shows that SegTransVAE outperforms previous methods in Dice Score and $95\%$-Haudorff Distance while having comparable inference time to a simple CNN-based architecture network. The source code is available at:\url{https://github.com/itruonghai/SegTransVAE}. 
\end{abstract}
\begin{keywords}
Transformer, Variational Autoencoder, Medical Image Segmentation, MRI brain tumor, CT kidney.
\end{keywords}
\section{Introduction}
\label{sec:intro}

Since the introduction of U-Net \cite{b1}, many state-of-the-art deep neural networks for medical image segmentation have been proposed. CNN-based segmentation networks such as U-Net \cite{b1}, and SegresnetVAE \cite{b2} are developed on a symmetric encoder-decoder architecture with skip connection, which combines high resolution features from the contracting path with the upsampled output. Then, this information can then be learned by a successive convolution layer to assemble a more precise output. However, they pose their limitation on learning global context and long-range spatial dependencies. As a result, this raises challenges to learn global semantic information which plays a critical role in segmentation tasks.

Transformer-based models in the natural language processing (NLP) domain have achieved state-of-the-art results. Inspired by attention mechanisms \cite{b3} in NLP, recent research such as UNETR \cite{b4} surpasses the aforementioned limitation in segmentation task by exploiting this mechanism. The self-attention mechanism in the transformers enables them to dynamically highlight the crucial features of sequences and learn their long-range dependencies. UNETR \cite{b4} leverages the power of transformers for volumetric medical image segmentation. A pure transformer is utilized as the encoder to learn contextual information from the embedded input patches. The extracted representations from the transformer encoder are merged with a decoder via skip connections at multiple resolutions to predict segmentation outputs. However, local structures are ignored when directly splitting images into patches as tokens for transformer, as mentioned in the research of Yuan et al. \cite{b5}. Moreover, UNETR \cite{b4} lacks inductive bias such as translation equivariance and locality, and therefore does not generalize well when trained on insufficient amounts of data.

In this work, a novel network named SegTransVAE is proposed to complement the drawbacks of existing studies. SegTransVAE is built upon an encoder-decoder architecture with the variational autoencoder (VAE) branch as the encoder regularization to the network to reconstruct the input images jointly with segmentation. Thanks to VAE branch, the proposed network can avoid the overfitting problem. First, the encoder of the network uses 3D CNN to extract the volumetric spatial features and downsample the input 3D images, which effectively captures the local 3D context information. Second, each volume is reshaped into a vector and fed into the transformer encoder for global feature modeling. Third, the 3D CNN decoder takes the feature embedding from transformer and performs progressive upsampling while the extracted representations from the encoder are concatenated with a decoder via skip connections at multiple resolutions to predict segmentation outputs.

\section{Method}
\label{sec:method}
\begin{figure*}[htb]
  \centering
  \includegraphics[width=0.85\textwidth]{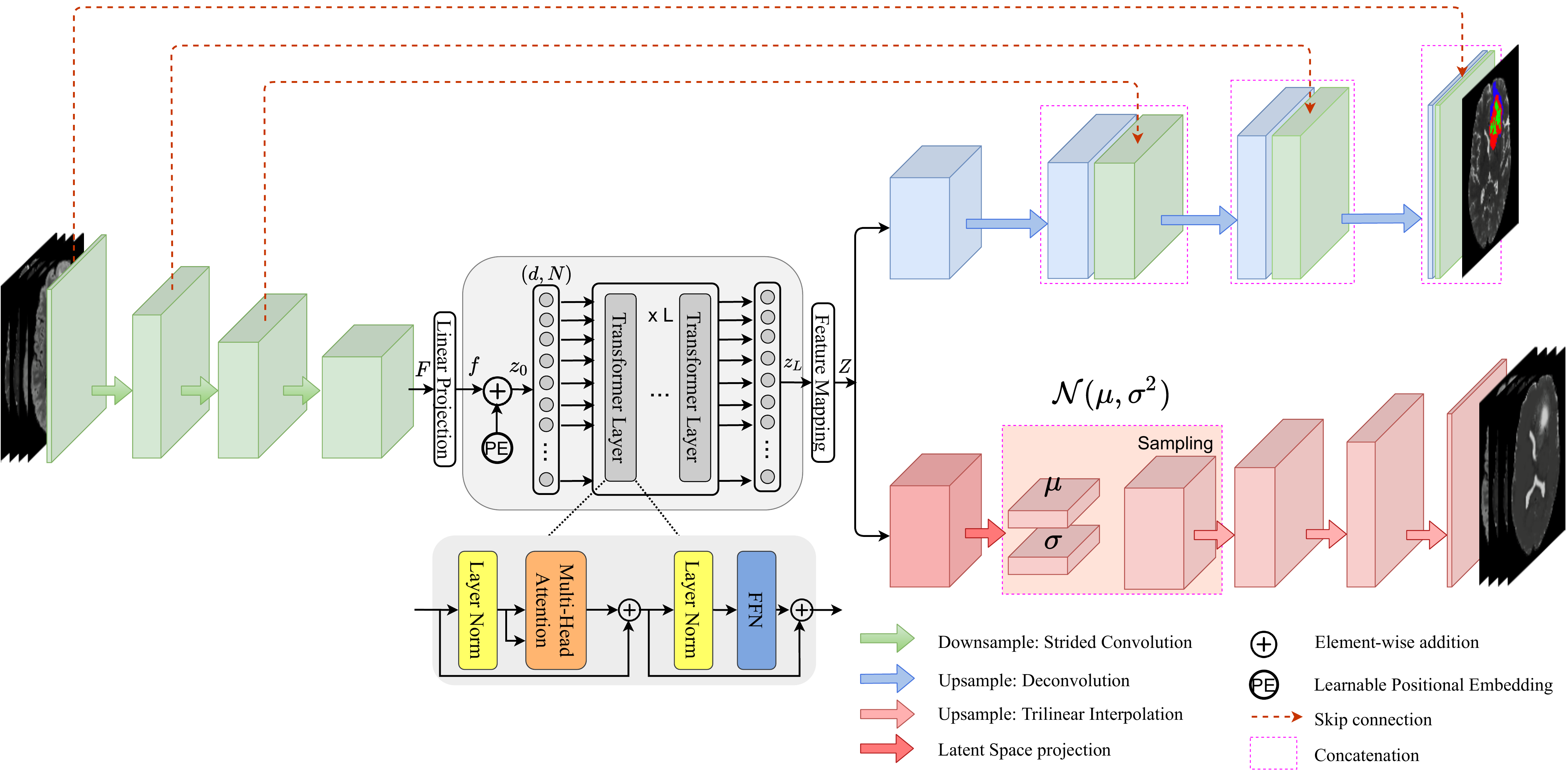}
  \caption{Overview architecture of proposed method.}
  \label{fig1}
\end{figure*}

The architecture of the proposed method is shown in Fig.\ref{fig1}. This approach follows encoder-decoder architecture with an asymmetrically larger encoder to extract image features, the transformer encoder to model the long-distance dependency in a global space and a smaller decoder to construct the segmentation mask. Also, an additional VAE branch is added to the endpoint of the transformer to reconstruct source images.   
\subsection{Encoder component}
\label{subsec:encoder}
Inspired by ResNet \cite{b6}, in this research, a modified-Resnet block is proposed in which consists of two convolutions with instance normalization \cite{ulyanov2016instance} and Leaky ReLU, followed by additive identity skip connection. This modified-ResNet block suffers from sparse gradients and shows a significant qualitative improvement. The encoder part uses the proposed modified-ResNet blocks. In order be able to model the image local context information across spatial and depth dimensions for volumetric segmentation, the modified-ResNet blocks are stacked with downsampling to gradually encode input $X \in \mathbb{R}^{C \times H \times W \times D}$ images into low-resolution and high-level feature representation $F \in \mathbb{R}^{K \times \frac{H}{8} \times \frac{W}{8} \times \frac{D}{8}}$. After that, this representation is fed into the transformer encoder to further learn long-range correlations with a global receptive field.
\subsection{Transformer component}
\label{subsec:transformer}
\subsubsection{Feature embedding}
\label{subsubsec:embedding}
A linear projection is used to project the feature map $F$ from $K$ dimensions to a $d$ dimensional embedding space $f$ in order to ensure a comprehensive representation of each volume. In order to encode the location information, the learnable position embeddings \cite{b7} are used and fused with the $d \times N$ feature map $f$ by direct addition, where $N = \frac{H}{8} \times \frac{W}{8} \times \frac{D}{8}$. This creates the feature embeddings as follows:
\begin{equation}
    \label{eq1}
    z_0 = f + PE = W \times F + PE,
\end{equation}
where the linear projection matrix is $W$, the position embeddings is $PE \in \mathbb{R}^{d \times N}$, and the feature embeddings is $z_0 \in \mathbb{R}^{d \times N}$.

\subsubsection{Transformer encoder}
\label{subsubsec:transfomer}
A stack of transformer layers \cite{b8} is utilized to construct transformer encoder in which each transformer layer consisting of Multi-Head Attention (MHA) and Feed Forward Network (FFN) sublayers according to
\begin{equation}
    \label{eq2}
    z'_l = \text{MHA(LN}(z_{l - 1})) + z_{l - 1},
\end{equation}
\begin{equation}
    \label{eq3}
    z_l = \text{FFN(LN}(z'_l)) + z'_l,
\end{equation}
where LN refers to the layer normalization and $z_l$ denotes the output of $l$-th transformer layer.
\subsubsection{Feature mapping}
\label{subsubsec:mapping}
A feature mapping module is added to project the sequence data back to a standard feature map. Then, this feature map is fed as the input dimension of 3D CNN decoder. In feature mapping module, the output sequence of transformer is $z_L \in \mathbb{R}^{d \times N}$ is first reshaped into $d \times \frac{H}{8} \times \frac{W}{8} \times \frac{D}{8}$ then a convolution block is employed to reduce the channel dimension from $d$ to $K$. Finally a feature map $Z \in \mathbb{R}^{K \times \frac{H}{8} \times \frac{W}{8} \times \frac{D}{8}}$ is obtained.

\subsection{Decoder component}
\label{subsec:decoder}
The encoder also uses modified-ResNet blocks to perform feature upsampling and pixel-level segmentation, but with a single block per spatial level. Each decoder level begins with an upsizing to reduce the number of features by a factor of 2 and double the spatial dimension, followed by a concatenation of encoder output of the equivalent spatial level. The end of the decoder has the same spatial size as the original image and the number of features equal to the initial input feature size, followed by $1\times1\times1$ convolution into 3 channels and a sigmoid function. 

\subsection{VAE component}
\label{subsec:vae}
Variational autoencoder (VAE) is added to reconstruct the volumetric input segmentation. The main role of VAE branch is to avoid the overfitting problem and to increase the network generalization. From the encoder endpoint output, the input is reduced to a lower-dimensional space of $256$ in which $128$ represents for mean, and the rest represents for standard deviation. A sample is drawn from the Gaussian distribution with the given mean and standard deviation $\mathcal{N} (\mu, \,\sigma^{2})$, then reconstructed into the input image dimensions following the same architecture as the decoder.  

\subsection{Loss Function}
Let $y$ and $\hat{y}$ be the ground truth of segmentation and the prediction of the model, respectively. To avoid training data having no label as $\hat{y} = y = 0$, $\varepsilon$ is added into numerator and denominator. Dice Loss is as follows
\begin{equation}
    \label{eq4}
    \L_{\text{Dice}}(y, \hat{y})= 1 - \frac{2\hat{y}y + \varepsilon}{\hat{y} + y + \varepsilon }.
\end{equation}


VAE loss is a total loss of reconstruction loss on VAE $\L_{\text{Rec}}$ branch and standard VAE penalty term $\L_{\text{KL}}$. Let $x_{\text{reconstruction}}$ and $x$ denote the reconstruction image and input image, respectively.

In this study, $\L_{\text{Rec}}$ is the mean square error over each voxels:
\begin{equation}
    \label{eq5}
    \L_{\text{Rec}}= \lVert x_{\text{reconstruction}}  - x \rVert^2_2.
\end{equation}

$\L_{\text{KL}}$ is a Kullback–Leibler divergence between the estimated normal distribution $ \mathcal{N} (\mu, \,\sigma^{2})$ and a prior distribution $\mathcal{N} (0, 1)$ as
\begin{equation}
    \label{eq6}
    \L_{\text{KL}} = \frac{1}{N_{\text{total voxels}}}\sum \mu^2 + \sigma^2 - \log \sigma^2 - 1,
\end{equation}
where $N_{\text{total voxels}}$ is the total number of image voxels.

The final loss function is the combination of Dice Loss and VAE Loss as follow
\begin{equation}
    \label{eq6}
    \L = \L_{\text{Dice}} + 0.1\times ( \L_{\text{Rec}} + \L_{\text{KL}}).
\end{equation}
A hyper-parameter (regularization factor weight) of 0.1 is chosen to provide a good balance between dice loss and VAE loss as \cite{b2}.
\section{Experiment}
\label{sec:experiment}
\subsection{Experimental Setup}
\label{subsec:setup}
\subsubsection{Dataset} 
The proposed method is evaluated on newly introduced BraTS 2021 \cite{b9} and KiTS19 \cite{b10}. BraTS 2021 \cite{b9} provides a 3D brain MRI dataset with tumor segmentation labels annotated. The training dataset comprises 1251 cases for training and 219 for validation rigidly aligned and resampled to a uniform isotropic resolution of $1mm^3$. The input image size is $240 \times 240 \times 155$. The KiTS19 \cite{b10} dataset is a collection of segmented CT imaging and treatment outcomes for 300 patients treated with partial or radical nephrectomy between 2010 and 2018.  Since the validation data of BraTS 2021 is private and it is not provided the ground truth, in this evaluation, 1251 cases is split as 1000 cases for training$/$validation and 251 cases for testing. Due to the small number of training images in KiTS19 \cite{b10}, five-fold cross-validation is chosen to evaluate proposed method and conventional models on this dataset. During training, the BraTS 2021 \cite{b9} input images are cropped of size $128 \times 128 \times 128$ while KiTS19 \cite{b10} input images are cropped of size $128 \times 160 \times 256$.
\subsubsection{Evaluation Metrics}
The metrics Dice score and $95\%$ - Hausdorff distance (HD) are used for quantitative evaluation. 

\subsection{Quantitative Results}
\label{subsec:results}
\subsubsection{BraTS 2021} 

 \begin{figure}[!htb]
    \begin{subfigure}[c]{0.09\textwidth}
        \centering
        \caption*{\textbf{Ground truth}}
        \includegraphics[width=0.9\linewidth]{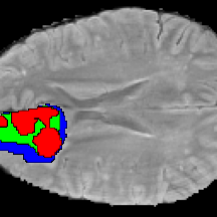}
    \end{subfigure} \hfill
    \begin{subfigure}[c]{0.09\textwidth}
        \centering
        \caption*{\textbf{SegTransVAE}}
        \includegraphics[width=0.9\linewidth]{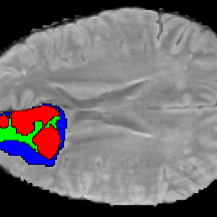}
    \end{subfigure} \hfill
    \begin{subfigure}[c]{0.09\textwidth}
        \centering
        \caption*{\textbf{SegresnetVAE}}
        \includegraphics[width=0.9\linewidth]{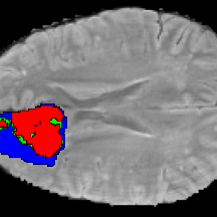}
    \end{subfigure} \hfill
    \begin{subfigure}[c]{0.09\textwidth}
        \centering
        \caption*{\textbf{3D U-Net}}
        \includegraphics[width=0.9\linewidth]{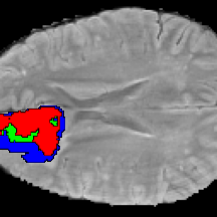}
    \end{subfigure} \hfill
    \begin{subfigure}[c]{0.09\textwidth}
        \centering
        \caption*{\textbf{UNETR}}
        \includegraphics[width=0.9\linewidth]{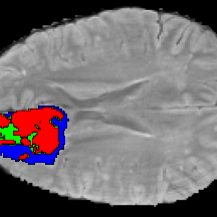}
    \end{subfigure}
    \caption{The visual comparison of BraTS segmentation results where red, green, blue are the enhancing tumor, core tumor and whole tumor, respectively.}
\label{fig2}
\end{figure}  

\begin{table}[!htb]
\centering
\caption{Dice Score and $95\%$-HD comparison on BraTS.}
\resizebox{0.45\textwidth}{!}{%
\begin{tabular}{|c|c|c|c|c|c|c|}
\hline
\multirow{2}{*}{Method} & \multicolumn{3}{c|}{Dice Score (\%)}             & \multicolumn{3}{c|}{95\%-HD (mm)}             \\ \cline{2-7} 
                        & \textbf{ET}    & \textbf{WT}    & \textbf{TC}    & \textbf{ET}   & \textbf{WT}   & \textbf{TC}   \\ \hline\hline
3D U-Net \cite{b13}     & 80.69 & 85.00 & 90.11 & 4.83 & 6.20 & 8.99 \\ \hline
UNETR  \cite{b4}      & 82.18 & 85.14 & 89.46 & 5.63 & 7.62 & 13.18 \\ \hline
SegresnetVAE \cite{b2} & 84.46 & 89.52 & 92.35 & 3.24 & 3.79 & 6.36 \\ \hline
\textbf{SegTransVAE}    & \textbf{85.48} & \textbf{90.52} & \textbf{92.60} & \textbf{2.89} & \textbf{3.57} & \textbf{5.84} \\ \hline
\end{tabular}%
}

\label{comparision:brats}
\end{table}
 In this experiment, the proposed method SegTransVAE is compared with state-of-the-art 3D approaches including 3D U-Net \cite{b13}, UNETR \cite{b4}, and SegresnetVAE \cite{b2}. Table \ref{comparision:brats} illustrates the Dice Score comparison between SegTransVAE and previous methods. It is clear that SegTransVAE outperforms previous research as it achieves the Dice Score of $85.48\%$, $90.42\%$, and $92.60\%$ on ET, WT and TC, respectively. In terms of $95\%$ - Hausdorff Distance, Table \ref{comparision:brats} shows that SegTransVAE also achieves considerable improvement. It is clear that due to leveraging CNN for high-level features extracting and transformer for global feature modeling, the proposed method shows its significant improvement in segmentation. It is obvious that in Fig. \ref{fig2} SegTransVAE creates segmentation masks of brain tumors more precisely and especially generates much better segmentation masks of the small tumor as enhancing tumor.
\subsubsection{KiTS 2019} 
\begin{figure}[!htb]    
    \begin{subfigure}[c]{0.09\textwidth}
        \centering
        \caption*{\textbf{Ground truth}}
        \includegraphics[width=0.9\linewidth]{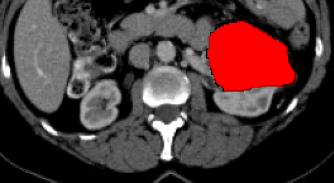}
    \end{subfigure} \hfill
    \begin{subfigure}[c]{0.09\textwidth}
        \centering
        \caption*{\textbf{SegTransVAE}}
        \includegraphics[width=0.9\linewidth]{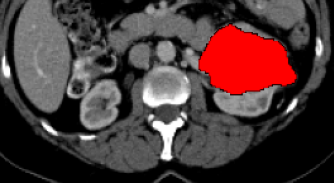}
    \end{subfigure} \hfill
    \begin{subfigure}[c]{0.09\textwidth}
        \centering
        \caption*{\textbf{SegresnetVAE}}
        \includegraphics[width=0.9\linewidth]{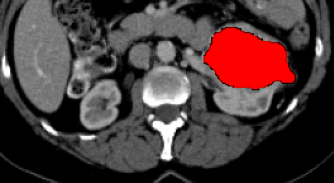}
    \end{subfigure} \hfill
    \begin{subfigure}[c]{0.09\textwidth}
        \centering
        \caption*{\textbf{3D U-Net}}
        \includegraphics[width=0.9\linewidth]{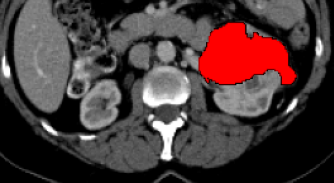}
    \end{subfigure} \hfill
    \begin{subfigure}[c]{0.09\textwidth}
        \centering
        \caption*{\textbf{UNETR}}
        \includegraphics[width=0.9\linewidth]{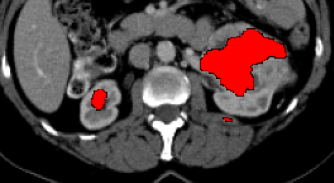}
    \end{subfigure}  
    
    \smallskip
    
    \begin{subfigure}[c]{0.09\textwidth}
        \centering
        \includegraphics[width=0.9\linewidth]{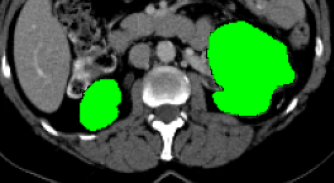}
    \end{subfigure} \hfill
    \begin{subfigure}[c]{0.09\textwidth}
        \centering
        \includegraphics[width=0.9\linewidth]{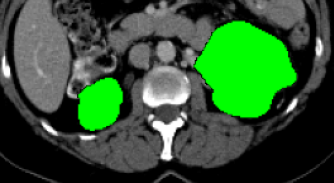}
    \end{subfigure} \hfill
    \begin{subfigure}[c]{0.09\textwidth}
        \centering
        \includegraphics[width=0.9\linewidth]{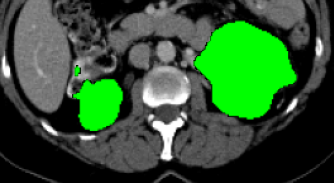}
    \end{subfigure} \hfill
    \begin{subfigure}[c]{0.09\textwidth}
        \centering
        \includegraphics[width=0.9\linewidth]{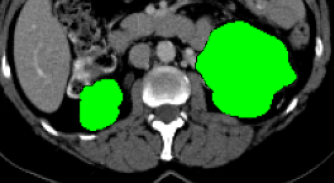}
    \end{subfigure} \hfill
    \begin{subfigure}[c]{0.09\textwidth}
        \centering
        \includegraphics[width=0.9\linewidth]{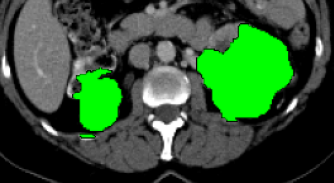}
    \end{subfigure}
    \caption{The visual comparison of KiTS segmentation results where red and green are tumor and kidney, respectively.}
\label{fig3}
\end{figure}
\begin{table}[!htb]
\centering
\caption{Dice Score and $95\%$-HD comparison of Kidney.}

\resizebox{0.45\textwidth}{!}{%
\begin{tabular}{|c|c|c|}
\hline
\multirow{2}{*}{Method} & \multicolumn{2}{c|}{\textbf{Kidney}}                          \\ \cline{2-3} 
                        & Dice Score (\%)           & 95\%-HD (mm)             \\ \hline\hline
3D U-Net   \cite{b13}   & 92.37 $\pm$ 4.54          & 6.32 $\pm$ 1.83          \\ \hline
UNETR      \cite{b4}    & 91.86 $\pm$ 1.29          & 8.84 $\pm$ 1.78          \\ \hline
SegresnetVAE  \cite{b2} & 94.86 $\pm$ 0.69          & 4.28 $\pm$ 0.97          \\ \hline
\textbf{SegTransVAE}    & \textbf{95.28 $\pm$ 0.85} & \textbf{3.28 $\pm$ 1.19} \\ \hline
\end{tabular}%
}
\label{comparision:kidney}
\end{table}
\begin{table}[!htb]
\centering
\caption{Dice Score and $95\%$-HD comparison of Tumor.}

\resizebox{0.45\textwidth}{!}{%
\begin{tabular}{|c|c|c|}
\hline
\multirow{2}{*}{Method} & \multicolumn{2}{c|}{\textbf{Tumor}}                            \\ \cline{2-3} 
                        & Dice Score (\%)           & 95\%-HD (mm)              \\ \hline\hline
3D U-Net   \cite{b13}   & 60.41 $\pm$ 4.14          & 42.02 $\pm$ 4.05          \\ \hline
UNETR      \cite{b4}    & 34.87 $\pm$ 3.80          & 60.43 $\pm$ 9.43          \\ \hline
SegresnetVAE  \cite{b2} & 63.67 $\pm$ 4.39          & 25.86 $\pm$ 3.24          \\ \hline
\textbf{SegTransVAE}    & \textbf{66.31 $\pm$ 4.41} & \textbf{24.61 $\pm$ 2.49} \\ \hline
\end{tabular}%
}
\label{comparision:tumor}
\end{table}
The proposed method is also evaluated on KiTS 2019 dataset \cite{b10}. Tables \ref{comparision:kidney} and \ref{comparision:tumor} illustrate that SegTransVAE outperforms in tumor segmentation and shows comparable results in kidney segmentation of the conventional methods as 3D U-Net \cite{b13}, UNETR \cite{b4}, and SegresnetVAE \cite{b2}. In addition, the proposed method shows better results in kidney and tumor in every fold of the experiment. By utilizing VAE, SegTransVAE shows its significant results in the little availability of training data as KiTS 2019 dataset \cite{b10}. It is clear that in Fig. \ref{fig3}, SegTransVAE shows better performance in segmentation tumor and kidney.   
\subsection{Complexity}

\begin{table}[!htb]
\centering
\caption{Comparision of number of parameters and averaged inference time.}

\resizebox{0.45\textwidth}{!}{%
\begin{tabular}{|c|c|c|c|}
\hline
Method      & \#Params (M)  & Inference Time (s) \\ \hline\hline
3D U-Net \cite{b13}   & 5.6         & 0.45               \\ \hline
UNETR   \cite{b4}     & 101.7        & 0.38               \\ \hline
SegresnetVAE \cite{b2} &  7.5                    &    0.55            \\ \hline

\textbf{SegTransVAE} & 44.7         & 0.45               \\ \hline
\end{tabular}%
}
\label{complexity}
\end{table}
The complexity of SegTransVAE is compared to other models in terms of the number of parameters and the averaged inference time. The benchmark is calculated based on the input size of $(4, 128, 128, 128)$. Table \ref{complexity} illustrates that SegTransVAE has 44.7M parameters as compared to 101.7M parameters of UNETR \cite{b4} which makes \cite{b4} hard to converge, especially with high-resolution input. As a result, the proposed method outperforms \cite{b4} at all evaluation metrics on BraTS 2021 and KiTS19 datasets. Although CNN-based segmentation methods as 3D U-Net and SegresnetVAE \cite{b2} have fewer parameters than UNETR \cite{b4} and SegTransVAE, the GFLOPs benchmarks of  CNN-based methods are more than UNETR and SegTransVAE, with the GFLOPs benchmarks of 3D U-Net and SegresnetVAE are more than 1000 GFLOPs while those of UNETR and SegTransVAE are 358.8 GFLOPs and 607.5 GFLOPs, respectively. As a consequence, SegTransVAE is less complex than the CNN-based network. Moreover, SegTransVAE has the second-lowest averaged inference time after UNETR and is comparable to simple CNN-based architecture like 3D U-Net. Also, SegTransVAE is 20\% faster than SegresnetVAE.
\section{Conclusion}
\label{sec:conclusion}
A novel network named SegTransVAE is presented with the goal of complement the disadvantages of existing studies and the little availability of training data. SegTransVAE is built upon encoder-decoder architecture with the variational autoencoder (VAE) branch to the network to reconstruct the input images jointly with segmentation. transformer is also used for global feature modeling. Experiments on two distinct datasets demonstrate the superiority of the proposed method when compared to state-of-the-art methods including 3D U-Net \cite{b13}, UNETR \cite{b4}, and SegresnetVAE \cite{b2}. 

\vfill
\pagebreak

\bibliographystyle{IEEEbib}
\bibliography{strings,refs}

\end{document}